\documentclass[sn-nature]{sn-jnl}

\usepackage{graphicx}%
\usepackage{multirow}%
\usepackage{acro}%
\usepackage{amsmath,amssymb,amsfonts}%
\usepackage{amsthm}%
\usepackage{mathrsfs}%
\usepackage[title]{appendix}%
\usepackage{xcolor}%
\usepackage{textcomp}%
\usepackage{manyfoot}%
\usepackage{booktabs}%
\usepackage{algorithm}%
\usepackage{listings}%
\usepackage{svg}
\usepackage{caption}
\usepackage{makecell}

\DeclareAcronym{api}{
  short = API ,
  long  = Application Programming Interface
}
\DeclareAcronym{ssh}{
  short = SSH ,
  long  = Secure Shell
}
\DeclareAcronym{pht}{
  short = PHT ,
  long  = Personal Health Train,
  plural = s
}
\DeclareAcronym{pasta}{
  short = PASTA-4-PHT ,
  long  = Pipeline for Automated Security and Technical Audits for the Personal Health Train
}
\DeclareAcronym{gdpr}{
  short = GDPR ,
  long  = General Data Protection Regulation
}
\DeclareAcronym{da}{
  short = DA ,
  long  = Distributed Analytics
}
\DeclareAcronym{fl}{
  short = FL ,
  long  = Federated Learning
}
\DeclareAcronym{ccpa}{
  short = CCPA ,
  long  = California Consumer Privacy Act
}
\DeclareAcronym{dpa}{
  short = DPA ,
  long  = Data Protection Act
}
\DeclareAcronym{dpia}{
  short = DPIA ,
  long  = Data Protection Impact Assessment
}
\DeclareAcronym{sast}{
  short = SAST ,
  long  = Static Application Security Testing
}
\DeclareAcronym{dast}{
  short = DAST ,
  long  = Dynamic Application Security Testing
}
\DeclareAcronym{fair4rs}{
  short = FAIR4RS ,
  long  = FAIR Principles for research software
}
\DeclareAcronym{cvss}{
  short = CVSS ,
  long  = Common Vulnerability Scoring System
}
\DeclareAcronym{cve}{
  short = CVE ,
  long  = Common Vulnerabilities and Exposures
}
\DeclareAcronym{nvd}{
  short = NVD ,
  long  = National Vulnerability Database
}
\DeclareAcronym{cwe}{
  short = CWE ,
  long  = Common Weakness Enumeration
}
\DeclareAcronym{dbfs}{
  short = DBfS ,
  long  = Docker Bench for Security
}
\DeclareAcronym{ci}{
  short = CI ,
  long  = Continuous Integration
}
\DeclareAcronym{cd}{
  short = CD ,
  long  = Continuous Deployment
}
\DeclareAcronym{cicd}{
  short = CI/CD ,
  long  = Continuous Integration/Continuous Deployment
}
\DeclareAcronym{vcs}{
  short = VCS ,
  long  = Version Control System
}
\DeclareAcronym{ast}{
  short = AST ,
  long  = Abstract Syntax Tree,
  plural = s
}
\DeclareAcronym{rda}{
  short = RDA ,
  long  = Research Data Alliance
}
\DeclareAcronym{fair}{
  short = FAIR ,
  long  = Findable; Accessible; Interoperable and Reusable
}
\DeclareAcronym{gl}{
  short = GL ,
  long  = Granularity Level
}
\DeclareAcronym{rx}{
  short = RX ,
  long  = Receive (in networking context)
}
\DeclareAcronym{tx}{
  short = TX ,
  long  = Transmit (in networking context)
}
\DeclareAcronym{aws}{
  short = AWS ,
  long  = Amazon Web Services
}
\DeclareAcronym{sql}{
  short = SQL ,
  long  = Structured Query Language
}
\DeclareAcronym{dos}{
  short = DoS ,
  long  = Denial-of-Service
}
\DeclareAcronym{nist}{
  short = NIST ,
  long  = National Institute of Standards and Technology
}
\DeclareAcronym{w3c}{
  short = W3C ,
  long  = World Wide Web Consortium
}

\begin{document}

\title[PASTA-4-PHT: A Pipeline for Automated Security and Technical Audits for the Personal Health Train]{PASTA-4-PHT: A Pipeline for Automated Security and Technical Audits for the Personal Health Train}


\author*[1]{\fnm{Sascha} \sur{Welten}}\email{welten@dbis.rwth-aachen.de}

\author[1]{\fnm{Karl} \sur{Kindermann}}\email{karl.kindermann@rwth-aachen.de}

\author[1]{\fnm{Ahmet} \sur{Polat}}\email{ahmet.polat1@rwth-aachen.de}
\author[1]{\fnm{Martin} \sur{Görz}}\email{martin.goerz@rwth-aachen.de}
\author[2]{\fnm{Maximilian} \sur{Jugl}}\email{Maximilian.Jugl@medizin.uni-leipzig.de}
\author[1]{\fnm{Laurenz} \sur{Neumann}}\email{laurenz.neumann@dbis.rwth-aachen.de}
\author[1]{\fnm{Alexander} \sur{Neumann}}\email{neumann@dbis.rwth-aachen.de}
\author[3]{\fnm{Johannes} \sur{Lohmöller}}\email{lohmoeller@comsys.rwth-aachen.de}
\author[3]{\fnm{Jan} \sur{Pennekamp}}\email{pennekamp@comsys.rwth-aachen.de}
\author[1]{\fnm{Stefan} \sur{Decker}}\email{decker@dbis.rwth-aachen.de}


\affil*[1]{\orgdiv{RWTH Aachen University}, \orgname{Chair of Computer Science 5}, \orgaddress{\street{Ahornstrasse 55}, \city{Aachen}, \postcode{52074}, \state{NRW}, \country{Germany}}}
\affil[2]{\orgdiv{Leipzig University Medical Center}, \orgname{Dept. Medical Data Science}, \orgaddress{\street{Stephanstraße 9c}, \city{Leipzig}, \postcode{04103}, \state{SN}, \country{Germany}}}
\affil[3]{\orgdiv{RWTH Aachen University}, \orgname{Communication and Distributed Systems}, \orgaddress{\street{Ahornstrasse 55}, \city{Aachen}, \postcode{52074}, \state{NRW}, \country{Germany}}}


\abstract{
\textbf{Background:} With the introduction of data protection regulations, the need for innovative privacy-preserving approaches to process and analyse sensitive data has become apparent.
One approach is the \ac{pht} that brings analysis code to the data and conducts the data processing at the data premises.
However, despite its demonstrated success in various studies, the execution of external code in sensitive environments, such as hospitals, introduces new research challenges because the interactions of the code with sensitive data are often incomprehensible and lack transparency.
These interactions raise concerns about potential effects on the data and increases the risk of data breaches.\\
\textbf{Results:} To address this issue, this work discusses a \ac{pht}-aligned security and audit pipeline inspired by DevSecOps principles, called \ac{pasta}.
The automated pipeline incorporates multiple phases that detect vulnerabilities, such as unintentionally or intentionally introduced weaknesses in the code of the \acs{pht}, before its deployment.
To thoroughly study its versatility, we evaluate \acs{pasta} in two ways.
First, we deliberately introduce vulnerabilities into a \acs{pht}.
Second, we apply our pipeline to five real-world \acsp{pht}, which have been utilised in real-world studies, to audit them for potential vulnerabilities.\\
\textbf{Conclusions:} Our evaluation demonstrates that our designed pipeline successfully identifies potential vulnerabilities and can be applied to real-world studies.
In compliance with the requirements of the \ac{gdpr} for data management, documentation, and protection, our automated approach supports researchers using the \ac{pht} in their data-intensive work and reduces manual overhead.
\acs{pasta} can be used as a decision-making tool to assess and document potential vulnerabilities in code for data processing.
The associated artefacts of this article, along with the pipeline configuration, are available online for adaptation and reuse.
Ultimately, our work contributes to an increased security and overall transparency of data processing activities within the \acs{pht} framework.
}

\keywords{Audits, Security, Code Evaluation, Personal Health Train, DevOps, DevSecOps}



\maketitle

\section{Background}
\label{sec:introduction}

Data privacy is a significant concern for many organisations and initiatives~\cite{Wirth2021,abouelmehdi2017bigdata,Gaye2014}.
Especially in healthcare, data privacy plays an important role due to the inherently sensitive nature of data instances that are collected, processed, and stored in highly secured IT infrastructures~\cite{Wirth2021,Gaye2014}.
To protect personal information and foster privacy-preserving processing, decision-makers have enacted regulations, such as the \ac{gdpr}\footnote{\ac{gdpr}: \url{https://gdpr-info.eu/} (accessed on 11.06.2024)} in the EU, \ac{ccpa}\footnote{\ac{ccpa}: \url{https://oag.ca.gov/privacy/ccpa} (accessed on 11.06.2024)} in California, and the \ac{dpa}\footnote{\ac{dpa}: \url{https://www.gov.uk/data-protection} (accessed on 11.06.2024)} in the UK~\cite{beyanDistributedAnalyticsSensitive2020,Choudhury2024,Wirth2021}.
These regulations impact the way how researchers access data for research purposes.

In recent years, the concepts of \ac{da} or \ac{fl} have emerged (see Figure~\ref{fig:vulnerabilities}) \cite{beyanDistributedAnalyticsSensitive2020,Wirth2021,Kim2024,Choudhury2024}.
These approaches move the analysis to the data rather than bringing the data to the analysis~\cite{beyanDistributedAnalyticsSensitive2020,weltenPrivacyPreservingDistributedAnalytics2022}.
This method enables accessing and analysing data while simultaneously complying with a wide range of data protection regulations~\cite{Kim2024}.
This paradigm shift has brought up several implementations, such as DataSHIELD or the \ac{pht} that follow the paradigm of analysis-to-data but differ in their technology stack and the way how the data analysis is sent to the data holders~\cite{Gaye2014,beyanDistributedAnalyticsSensitive2020,budin-ljosne_datashield_2014,weltenDams2021,Choudhury2024}.
However, even though the method of \ac{da} has shown promising results in numerous studies, its application presents new challenges and offers opportunities for further research~\cite{vanSoest2018Using,BoninoDaSilvaSantos2022,weltenPrivacyPreservingDistributedAnalytics2022,herr2022bringing,budin-ljosne_datashield_2014}.

\subsection{Problems and Challenges in Distributed Analytics}

\begin{figure}[htbp]
\centering
    \includegraphics[width=0.99\linewidth]{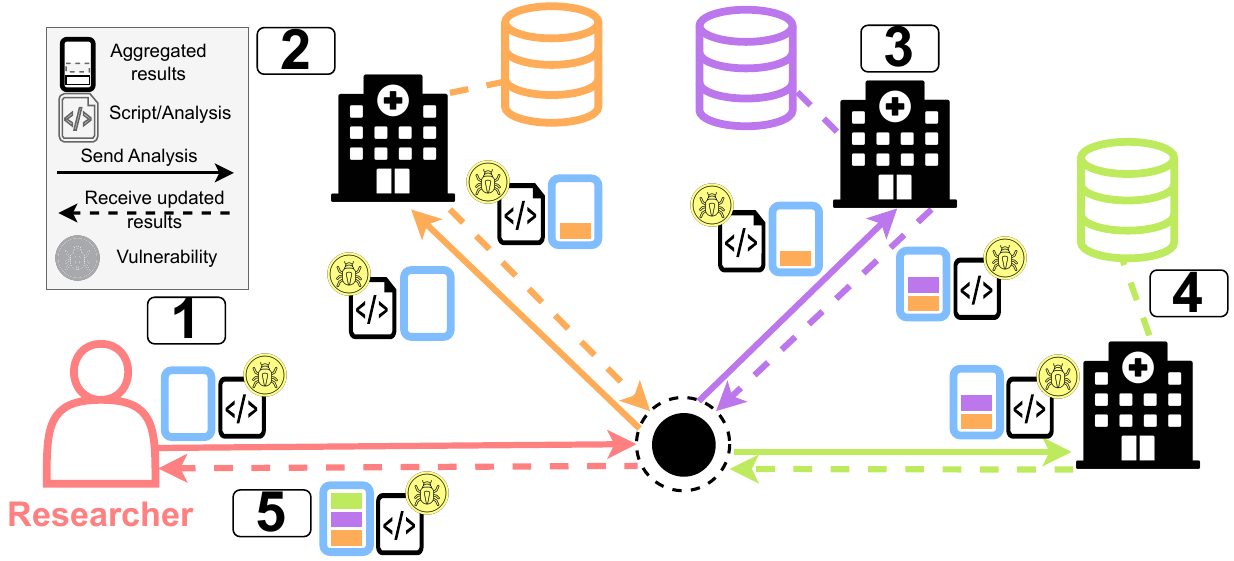}
    \caption{Workflow of \acf{da} with multiple data holders such as hospitals. Analysis results are aggregated after each execution at the respective hospitals and returned to the analysis requester. If the code is not properly reviewed, it may contain vulnerabilities that could compromise the IT of the sensitive environments.}
    \label{fig:vulnerabilities}
\end{figure}

The \ac{da} scenario reveals potential threats to the infrastructure of the data holders or the data subjects.
For example, introducing external code (the analysis) into secure IT infrastructures raises concerns, as this code interacts with potentially confidential data (see Figure~\ref{fig:vulnerabilities}).

At this level, the algorithms operate as blackboxes, and the question arises of how the analysis processes the data and what information leaves the institutional borders.
This lack of transparency in the operation of the analysis code poses considerable risks to the data-holding institution, creating opportunities to exploit (undetected) vulnerabilities in the analysis code.
Inspired by the \ac{nist}, we define a vulnerability as any weakness, flaw, or intentionally introduced malicious code in the data analysis (or its associated software components) that could potentially be exploited by external attackers~\cite{dempsey_automation_2020}.
There might be several ways to introduce vulnerabilities into the analysis code and its components.
While every participant (e.g. researchers) in a \ac{da}-enabling infrastructure is generally expected to be trustworthy, unintentional vulnerabilities introduced by the researchers can potentially be exploited by third parties or external actors.
The more realistic concern is that malicious actors could actively add malicious code to the data analysis code.
The consequence is that without code reviews, such exploitation could lead to data breaches or compromised IT infrastructures~\cite{Cheng2017,Elahi2010,Brilhante2024}.
Table~\ref{tab:code_threats} provides some examples of potential threats that may arise during data analyses.

Beyond the aspect of security, the described lack of transparency further complicates the compliance to the obligatory data management and protection measures, such as the documentation and monitoring of the data processing activities prescribed by the \ac{gdpr}~\cite{voigt2017eu}.

\begin{table}[htbp]
\centering
\caption{Some examplary and code-based threats in the analysis code. For each potential threat, we developed an example scenario where the threat could occur. Note that some threats may work in combination, and the scenarios might overlap.}
\label{tab:code_threats}
\begin{tabular}{|l|l|}
\toprule
\textbf{Threat} & \textbf{Scenario} \\
\midrule
Data Exfiltration & Malware copies data to an external server. \\
\hline
Code Injection & Inject code to compromise systems, e.g. the database. \\
\hline
Unauthorised Execution & Run unauthorised commands, access sensitive data. \\
\hline
Supply Chain Attack & Insert malicious code, compromise libraries. \\
\hline
Data Poisoning & Modify data in the data source. \\
\hline
Misconfiguration & Use weak security settings to gain access to the IT infrastructure. \\
\bottomrule
\end{tabular}
\end{table}

These current challenges in \ac{da} are not new, and some policies have already been implemented to proactively address the detection of vulnerabilities~\cite{Wirth2021,budin-ljosne_datashield_2014}.
For example, DataSHIELD established a process to validate the remote commands for the data analysis, as detailed by Budin-Ljøsne et al.~\cite{budin-ljosne_datashield_2014}.
Each command is (manually) reviewed before being included in a new software release to reduce the risk of potential data disclosures~\cite{budin-ljosne_datashield_2014}.
Nevertheless, to the best of our knowledge, other approaches, such as the \ac{pht}, lack such an audit process and protection guarantees for the analysis code to reveal potential undetected vulnerabilities, which may limit its broader acceptance~\cite{Wirth2021}.

\subsection{The Personal Health Train}

Essentially, the concept of the \ac{pht} has been introduced to promote the \ac{fair} Principles, or their adaptation known as \ac{fair4rs}, which are relevant to the work discussed here~\cite{weltenPrivacyPreservingDistributedAnalytics2022,beyanDistributedAnalyticsSensitive2020,herr2022bringing,RDAFORCE112020CodeID, ChueHong2022FAIR4RS}.
There have been several infrastructures developed that adapted the \ac{pht} approach and applied it to various data analysis use cases already~\cite{vanSoest2018Using,BoninoDaSilvaSantos2022,weltenPrivacyPreservingDistributedAnalytics2022,herr2022bringing}.
Field of applications are, for example, COVID-19, Cancer, Radiomics, or Bioinformatics~\cite{BoninoDaSilvaSantos2022,Zhang2023Secure,Shi2019Distributed,herr2022bringing}.
As described by Beyan et al., the \ac{pht} is modelled after the analogy of a Train visiting Stations on Tracks (see Figure~\ref{fig:pht})~\cite{beyanDistributedAnalyticsSensitive2020}:

\begin{figure}[htbp]
\centering
    \includegraphics[width=0.95\linewidth]{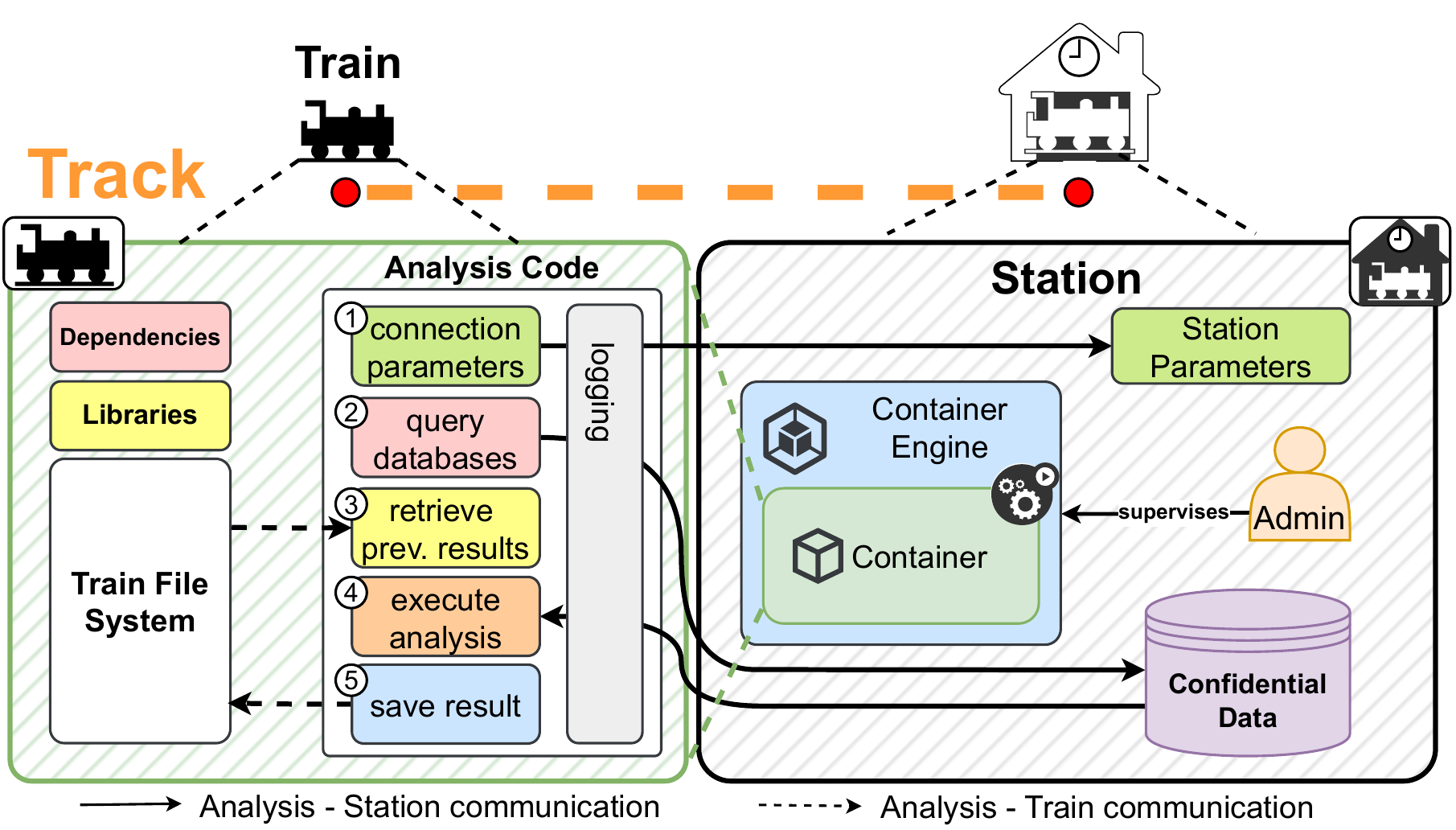}
    \caption{The \acf{pht} infrastructure with Container Trains. A Container Train is based on software containers and contains the analysis code, dependencies, libraries, and results. At each Station, the Train is executed, allowing access to and processing of data during the analysis execution (Steps 1--5). We assume that the analysis code, along with its dependencies and libraries (left), might contain vulnerabilities.}
    \label{fig:pht}
\end{figure}

\textbf{The Stations} are endpoints, e.g. at a hospital, which hold the confidential data and facilitate the execution of the analysis (see Figure \ref{fig:pht})~\cite{weltenPrivacyPreservingDistributedAnalytics2022}.
An admin runs the arrived Train and inspects the analysis results stored in the Train after its execution~\cite{weltenPrivacyPreservingDistributedAnalytics2022}.
As defined but not further described by Beyan et al., the Train is executed within a secure environment separate from the data source, providing an initial layer of security~\cite{beyanDistributedAnalyticsSensitive2020}.
Thus, this setup gives the Station full control over its data and the returned results.
Finally, the Train creators (e.g. the researchers) can inspect the result of their analysis.

\textbf{The Tracks} are connections between Stations.
Tracks facilitate communication by receiving Trains from the developer and directing them towards selected Stations on a given route~\cite{beyan2020pht}.

\begin{figure}[htbp]
\centering
    \includegraphics[width=0.95\linewidth]{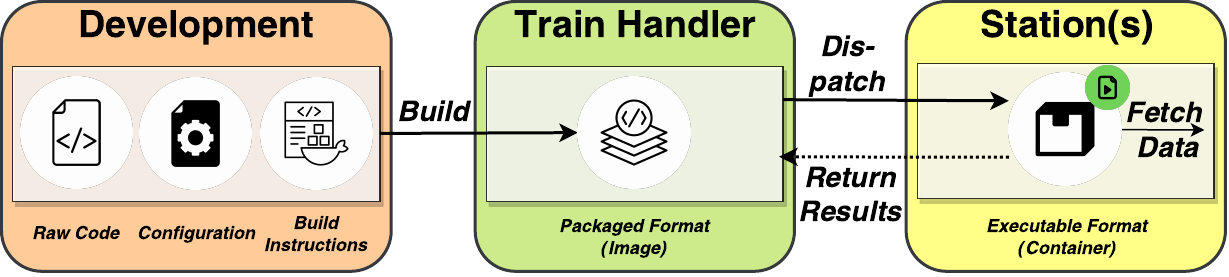}
    \caption{Train lifecycle diagram for Container Trains. First, a Train is represented in raw format. After a build step, the Train is represented by an image and managed by a Train handler (i.e.\ the \ac{pht} infrastructure) that orchestrates the Train images to the Stations. At a Station, the Train code needs to be executed. Hence, the image is transformed into an executable Container. Terminology partially adapted from Bonino et al.\ and Beyan et al.~\cite{BoninoDaSilvaSantos2022,beyanDistributedAnalyticsSensitive2020}.}
    \label{fig:containertrainlifecycle}
\end{figure}

\textbf{The Train}, in its most generic form, symbolises code or algorithms that are designed for interacting with data at Stations~\cite{BoninoDaSilvaSantos2022,beyan2020pht}.
The way how a Train interacts with data depends on the methodology the Train implements~\cite{BoninoDaSilvaSantos2022}.
Bonino et al. set up a taxonomy that categorises various types of Trains~\cite{BoninoDaSilvaSantos2022,beyan2020pht}.
This includes Query Trains, Message Trains, \ac{api} Trains, Script Trains, and, notably for this study, Container Trains.
Container Trains (see Figure~\ref{fig:pht}) encapsulate the analysis, libraries, and other dependencies in a container image~\cite{BoninoDaSilvaSantos2022, weltenPrivacyPreservingDistributedAnalytics2022,herr2022bringing}.
Each Train has a lifecycle, depicted in Figure~\ref{fig:containertrainlifecycle}, that we derived from the definitions by Bonino et al~\cite{BoninoDaSilvaSantos2022}.
The lifecycle consists of three stages within the development process of a Container Train: The initial coding and configuration phase, the construction of the Train into a packaged software image format, and the final execution stage, where the packaged code is transported to the Stations.
After a Train is dispatched to a Station, the software image is converted into a container at the Station, which then fetches the data.
At this point, it can interact with data by connecting to the data source, querying data, loading previous results, performing data analysis, and saving the results (Steps 1--5 in Figure~\ref{fig:pht}).
For the usage of Container Trains, a containerisation technology is needed, such as Docker\footnote{Docker: \url{https://www.docker.com} (accessed on 11.06.2024)}~\cite{vanSoest2018Using,BoninoDaSilvaSantos2022}.

At this point, it becomes apparent why the (Container) Trains can pose threats to the Stations.
By examining the structure of Container Trains (see Figure~\ref{fig:pht}) and their lifecycle (Figure~\ref{fig:containertrainlifecycle}), we can identify several potential attack surfaces, concerning raw, packaged or executable code, where malicious actors could compromise the \ac{pht} if vulnerabilities remain undetected.

\subsection{Focus of this Software Article}

In this paper, our primary objective is to identify and analyse potential sources of vulnerabilities in the \ac{pht}.
We first examine the specific stages of a Train's lifecycle (Figure~\ref{fig:containertrainlifecycle}) where vulnerabilities are likely to occur and develop strategies to better identify and manage these vulnerabilities.
Based on this foundation, we present a DevSecOps-inspired reference implementation called \ac{pasta}.
This pipeline is designed to identify and document vulnerabilities in Train-relevant assets, serving as a decision-support tool to determine whether a Train is secure.
Instead of a manual review of the analysis code (like in DataSHIELD), \ac{pasta} addresses the previously identified gap in audit solutions within the \ac{pht} by offering an automated and systematic approach to code review and vulnerability detection.
The tool includes both existing software and entirely new components, which are integrated into a holistic solution that is available as open-source software.
Furthermore, it can be modified for additional uses, such as extensions for broader audit perspectives~\cite{karlkindermann_saweltpasta-4-pht_2024}.
In the following section, we will describe our methodology and the (reference) implementation.
We further demonstrate its readiness for practical use across a variety of (research) settings.

\section{Implementation}\label{sec:background}

Ideally, the (sample) threats mentioned in Table~\ref{tab:code_threats} should be detected as early as possible in the process of software development - or, in our case, the development of the analysis.
To achieve this early detection, we begin by exploring best practices and relevant approaches that inspired the development of our audit pipeline (Section~\ref{subsection:methodolody}). 
Based on the potential sources of vulnerability (Section~\ref{subsec:requirementanalysis}), we present the design and implementation details of \ac{pasta} in Section~\ref{subsec:conceptualisation}.

\subsection{Methodology}
\label{subsection:methodolody}

We will first frame our scenario and present assumptions made to design our software (see Section~\ref{subsubsection:assumptions}).
Then, we examine DevOps (see Section~\ref{subsubsection:devopsanddevsecops}) and its extension, DevSecOps, as well as explore current methods for detecting vulnerabilities (Section~\ref{subsection:vuldetection}).
Note that we use the term Train interchangeably with Container Trains.

\subsubsection{Assumptions about the Scenario}
\label{subsubsection:assumptions}

First of all, the objective of our software is to detect vulnerabilities introduced in Container Trains.
This means we disregard any potential infrastructure-related threats, such as Man-in-the-Middle attacks.
As mentioned, we assume all participants in the infrastructure are trustworthy.
However, we expect two main ways vulnerabilities can be introduced into the Train code: accidentally, through the use of vulnerable software components, or intentionally, by malicious third parties who add malicious code into the Train before deployment.
We further assume that our audit pipeline is operated by the \ac{pht} infrastructure hosts and is thus (ideally) protected against manipulation by external attackers.
We interpret the audit pipeline as a pre-deployment step to detect vulnerabilities before the Train is deployed in the \ac{pht} infrastructure.
Hence, the pipeline represents a safeguard or portal to the \ac{pht} infrastructure.
Therefore, our software (or its components) is designed to support hosts that run and maintain the pipeline, and methods to bypass the pipeline are beyond the scope of this study.
Lastly, the goal is to make vulnerabilities visible rather than providing countermeasures.
The pipeline should leave the final decision about deploying the Train (accept or reject) to the infrastructure host.
Although this involves a final manual step, like DataSHIELD, our pipeline aims to be fully automated to avoid manual code reviews.

To achieve this, we will discuss the state-of-the-art practices in software development that inspired the development of the pipeline in the next section.

\subsubsection{DevOps and DevSecOps}
\label{subsubsection:devopsanddevsecops}

The Train lifecycle and its development workflow presented in Figure~\ref{fig:containertrainlifecycle} show similarities to processes observed in DevOps~\cite{devops,lwakatare2015dimensions}.
DevOps is an agile software development methodology that encompasses both the development (Dev) and operation (Ops) aspects of software~\cite{lwakatare2015dimensions}.
It especially focuses on integrating and automating software development processes~\cite{devops}.
This approach aims to shorten the development lifecycle, increase deployment frequency, and create more dependable releases~\cite{devops,lwakatare2015dimensions}.
One of the core characteristics of DevOps is the automation of software integration and delivery through so-called \ac{cicd} pipelines~\cite{devops}.
While \ac{ci} refers to the process of automatically integrating code from multiple developers into a single software version, \ac{cd} involves consistently deploying new software versions to production~\cite{myrbakken2017devsecops}.
In connection with \ac{cicd} pipelines, virtualisation tools such as containerisation play an integral role in build and release processes~\cite{challengesindevsecops,combe2016docker}.
The default DevOps workflow can be complemented by a Security (Sec) component, yielding DevSecOps~\cite{myrbakken2017devsecops,challengesindevsecops}.
DevSecOps extends DevOps with software security methods and quality assurance~\cite{myrbakken2017devsecops,challengesindevsecops}.
Like DevOps, DevSecOps implements automated methods for ongoing and continuous auditing of software~\cite{myrbakken2017devsecops,challengesindevsecops}.
Given the relevance of software audits in this study, we will shortly introduce various methods, techniques, and tools in the following section.

\subsubsection{Vulnerability Detection and Security Audits}
\label{subsection:vuldetection}

While all vulnerability detection techniques share the same goal of discovering potential vulnerabilities in the code, they differ in terms of methodology and procedure~\cite{challengesindevsecops}.
There is the \textit{static} or the \textit{dynamic} analysis of the code, also referred to as \ac{sast} or \ac{dast}~\cite{challengesindevsecops,pan2019SAST,mateo2020SASTDAST,felderer2016security}.
\ac{sast} validates the code without executing it, aiming to identify vulnerabilities by examining the source code and its dependencies (whitebox testing).
On the other hand, \ac{dast} executes the code to detect vulnerabilities by observing its behaviour during runtime (blackbox testing) ~\cite{pan2019SAST}.

Typically, vulnerabilities are evaluated using the \ac{cvss}, which categorises the severity of vulnerabilities into standardised classifications~\cite{scarfone2009CVSS,wist2021vulDocker,Brilhante2024}.
\ac{cvss} considers the ease of exploitation and the potential impact of a vulnerability and weighs them into a score, which is used to provide a general vulnerability rating~\cite{scarfone2009CVSS}.
In addition to this classification, vulnerability knowledge databases such as \ac{cve}\footnote{\ac{cve}: \url{https://www.cve.org/About/Overview} (accessed on 11.06.2024)} and \ac{nvd}\footnote{\ac{nvd}: \url{https://nvd.nist.gov} (accessed on 11.06.2024)} have been created to promote a shared terminology and comprehension of vulnerabilities by assigning unique identifiers to these vulnerabilities~\cite{combe2016docker,wist2021vulDocker}.
In contrast, the \ac{cwe}\footnote{\ac{cwe}: \url{https://cwe.mitre.org/about/index.html} (accessed on 11.06.2024)} catalogues the underlying flaws that can potentially result in vulnerabilities~\cite{wist2021vulDocker}.

The process of detecting vulnerabilities is often referred to as \textit{security auditing} that systematically validates the code against known security issues from, e.g. these databases~\cite{Stouffer2023,yamaguchi2012SAST}.
The results of such audits are usually documented in potentially standardised report documents, which serve as the inspiration for this work\footnote{Examples: \url{https://github.com/juliocesarfort/public-pentesting-reports} (accessed on 11.06.2024)}.
As previously stated, the methods and tools used for detecting vulnerabilities depend on the format of the software~\cite{mateo2020SASTDAST,felderer2016security}.
In addition to raw code, packaged formats, such as software images or containers, also exist (see Figure~\ref{fig:containertrainlifecycle}).

Software images are more complex than basic code bases as they encapsulate additional software fragments, which can also introduce unwanted vulnerabilities~\cite{combe2016docker}.
This can be prevented by using trusted images, which are images that are verified and published by official image stores such as Docker Hub\footnote{Docker Hub: \url{www.dockerhub.de} (accessed on 11.06.2024)} or other trusted third parties~\cite{dahlmanns2023secretsindocker}.
Yet, images get updates and new versions are released, which again can potentially introduce new vulnerabilities~\cite{combe2016docker,dahlmanns2023secretsindocker}.
For this reason, tools like Clair\footnote{Clair: \url{https://github.com/quay/clair} (accessed on 11.06.2024)}, grype\footnote{Grype: \url{https://github.com/anchore/grype} (accessed on 11.06.2024)}, trivy\footnote{Trivy: \url{https://github.com/aquasecurity/trivy} (accessed on 11.06.2024)}, Snyk\footnote{Snyk: \url{https://snyk.io} (accessed on 11.06.2024)}, Docker Scout\footnote{Docker Scout: \url{https://docs.docker.com/reference/cli/docker/scout/} (accessed on 11.06.2024)}, or Harbor\footnote{Harbor: \url{https://goharbor.io} (accessed on 11.06.2024)} are used for scanning images for vulnerabilities.
These tools mainly perform static analysis, but \ac{dast} is also necessary as containers in their executable state can also have vulnerabilities.
For example, tools exist to assist developers in running containers in a safer way by providing hints and warnings.
One such tool is the \ac{dbfs}\footnote{\ac{dbfs}: \url{https://github.com/docker/docker-bench-security} (accessed on 11.06.2024)}.
\ac{dbfs} informs the developer by checking configurations and creating a report about whether they contradict Docker's best practices, especially for security.
Another tool representative for dynamic analysis is Aqua Security's Dynamic Threat Analysis\footnote{DTA: \url{https://www.aquasec.com/products/container-analysis/} (accessed on 11.06.2024)}.
Their tool detects suspicious activities of the container, such as reading confidential credentials, code injection backdoors, or network traffic.

As presented, there are various approaches and tools available for Train developers or \ac{pht} infrastructure administrators that may be used to scan and detect vulnerable code.
Such security audits, like those integrated into a DevSecOps pipeline, can be an effective method for this detection.
In the following sections, we will present our implementation for a DevSecOps-inspired audit pipeline tailored specifically for applications within the \ac{pht}.
For this, we first explore possible sources of vulnerabilities in Container Trains (Section~\ref{subsec:requirementanalysis}).

\subsection{Determining the Sources of Vulnerabilities}
\label{subsec:requirementanalysis}

In line with our definition of a vulnerability in Section~\ref{sec:introduction}, a Train may include undesirable code leading to security or privacy problems or causing issues like system crashes or excessive use of resources that might strain the Station.
Because vulnerabilities can affect different areas and formats of software, as we have seen in Section~\ref{sec:background}, we must examine in what formats a Train can occur to design the pipeline.
Hence, we need to understand the evolution of the Train (code).
In Section~\ref{sec:background}, we have introduced the Train's lifecycle in Figure~\ref{fig:containertrainlifecycle} that reflects the Train's evolution from its initial development phase to its actual execution.
Building upon this conventional workflow in Train development shown in Figure~\ref{fig:containertrainlifecycle}, we have identified and elaborated on three key states that a Train progresses through its lifecycle, which we call \emph{Aggregation States}:

\begin{enumerate}
    \item \textbf{Aggregation State 1: \textit{Source Code.}} At this level, the focus is on the code base in its most fundamental form. This encompasses, e.g. the source code and configuration files, representing the Train in its raw and unaltered state. The code is not yet tailored for execution in any specific runtime environment.
    \item \textbf{Aggregation State 2: \textit{Packaged Code.}} In this phase, the static code is transformed into a (Train) image. This process involves encapsulating the code and its dependencies, environment settings, libraries, and other necessary components. The result of this process is, e.g. a transferable software image.
    \item \textbf{Aggregation State 3: \textit{Executable Code.}} Before a Train can be executed at a Station, the Train images must be transformed into an executable (Train) container. This instance represents the execution state of our Train and operates based on the packaged dependencies and the internal code.
\end{enumerate}

Each Aggregation State builds on top of the previous one and has specific code-related characteristics.
For vulnerabilities in state 1, we need to inspect the static code to detect security flaws without executing it.
Some vulnerabilities can include coding errors, insecure coding practices, or hard-coded credentials.
Hence, we need to analyse the code from a structural and syntactical perspective.
Further, moving one layer up and from a packaged perspective, vulnerabilities can arise due to insecure or outdated dependencies and libraries included in packages.
The encapsulation process, which transforms the code into an image, introduces another layer of complexity as new software is combined with each other to build the (Train) image.
These issues might not be visible until the code is in a more complex and integrated state.
Lastly, in Aggregation State 3, vulnerabilities on this level might only manifest when the code is executed.
Tools for this state cover runtime issues and services that might expose security flaws or misconfigurations.

With this three-layered approach in mind, we will present layer-specific methods for vulnerability detection that constitute \ac{pasta} in the following section.

\subsection{Implementation of \ac{pasta}}
\label{subsec:conceptualisation}

\begin{figure}[htbp]
    \centering
    \includegraphics[width=\linewidth]{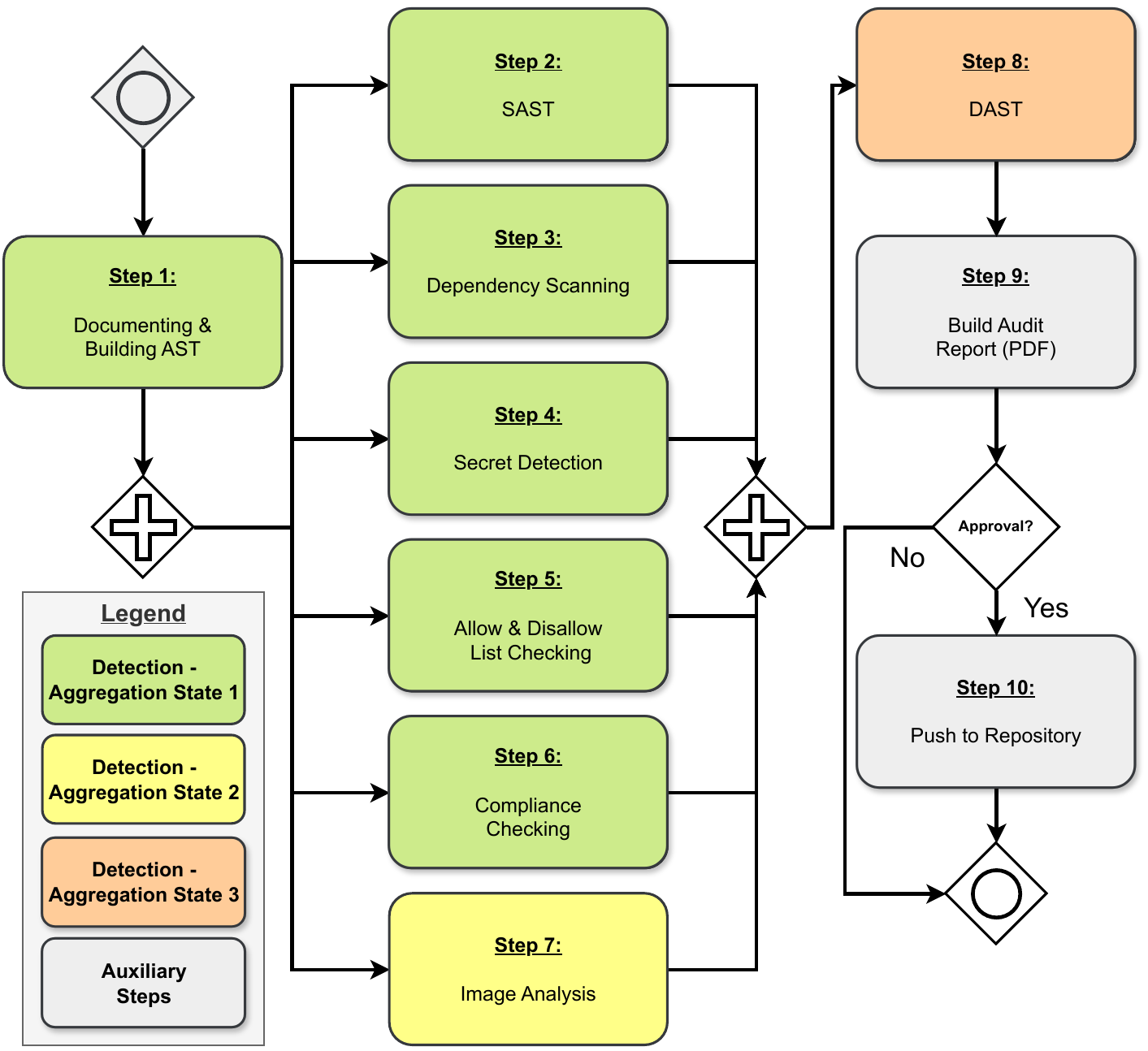}
    \caption{Formal visualisation of \ac{pasta}.}
    \label{fig:formalPipeline}
\end{figure}

Drawing inspiration from DevSecOps principles and established tools for software auditing, we present a range of detection mechanisms for each of the Aggregation States previously discussed (Section~\ref{subsec:requirementanalysis}).
Additionally, we shortly outline their focus and significance within the context of the \ac{pht} and, finally, demonstrate how we implemented them in \ac{pasta}.
A formal representation of our audit pipeline is given in Figure~\ref{fig:formalPipeline}, and the corresponding implementation details can be found in Figure~\ref{fig:pasta4pht}.
For each software component, we refer to the steps given in Figure~\ref{fig:formalPipeline}.
For more details about the implementation of the pipeline components, we refer to the supplementary materials or the corresponding references~\cite{karlkindermann_saweltpasta-4-pht_2024}.
For the sake of simplicity in this study, we consider the analysis code to be written in Python\footnote{Python: \url{https://www.python.org} (accessed on 11.06.2024)} as a commonly used language in data science.
However, the software we reuse in our pipeline is also compatible with other languages.

\subsubsection{Detection in Aggregation State 1 (Source Code)}
\label{subsubsection:countermeasuresaggrstate1}
Initially, we address the first state of a Train, which involves the raw analysis code.
The primary objective is to identify any vulnerabilities, coding mistakes, or security weaknesses present in the source code.
As initially described, the \ac{pht} often lacks transparency with respect to the code.
Due to this reason, we first describe our concept of code documentation to tackle this challenge.

\paragraph{Documentation of the Code (Step 1)}

In general, the documentation of code contributes to transparency and reproducibility, which in turn fosters trust~\cite{karthik2013versioning}.
In other words, if the Train code is made accessible to, e.g. the Station administrators, they are able to inspect the Train code.
Consequently, our approach is to establish a central repository for Train code, like a \ac{vcs}, to monitor changes and track the origin of changes.

\begin{figure}[htpb]
    \includegraphics[width=\linewidth]{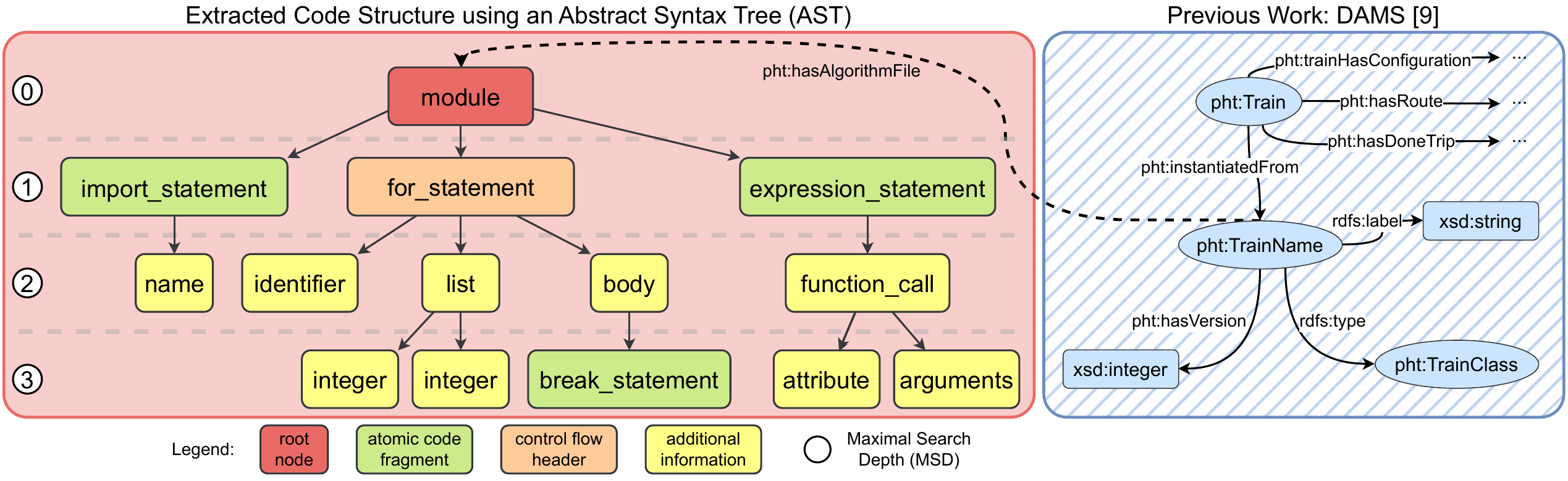}
    \caption{Integrating Train Code into a semantic framework using an \acf{ast}. We first extract the syntactical structure of the Train code by using \acsp{ast}. Given that \acsp{ast} are inherently a graph, we seamlessly merge this \ac{ast}-graph into a broader semantic context and enrich it with metadata~\cite{weltenDams2021}. For the sake of simplicity, we adopt the terminology for tree nodes from the \acs{ast} library (Tree-Sitter) and omitted the description of the arcs in this figure.}
    \label{fig:combined}
\end{figure}

To improve understanding of the code's structure and enable more sophisticated analyses, such as meta-analysis, in the future, our additional goal is two-fold.
First, we perform an extraction of the code's structure, and, secondly, we enrich it with additional metadata by incorporating it into a broader semantic context.
Especially, semantic mapping has several advantages as it allows for the association and linkage of specific code segments with known vulnerabilities, as catalogued in one of the vulnerability databases referred to in Section~\ref{sec:background}.
One common way to extract the structure of code is the utilisation of so-called \acp{ast}, which brings the Train code into a tree-shaped structure (see Figure~\ref{fig:combined})~\cite{neamtiu2005AST}.
This approach is beneficial in two ways.
First, the extracted syntactical structure of the code (the \ac{ast}) can be further reused for additional purposes such as code comparisons~\cite{neamtiu2005AST}.
Second, the code transformation into an \ac{ast} format aligns well with common Semantic Web technologies, which are predominantly graph-oriented~\cite{weltenDams2021}.
As a result, the code structure naturally takes the form of a graph, which can be integrated into a larger semantic framework.
In our specific context, the \ac{ast} is linked to an additional semantic graph, which supplements the Train code with extra metadata, such as details about the Train itself (e.g. name, origin).
Such Train metadata (schema) has already been introduced by the previous work of Welten et al.~\cite{weltenDams2021}.

\textbf{Implementation.}
Our process begins with the code base developed by a Train developer (see Step 1 in Figure~\ref{fig:formalPipeline}).
The starting point for our pipeline is GitLab as \ac{vcs}, to which the code is uploaded.
Beyond the versioning functionality, GitLab facilitates the use of the already introduced \ac{cicd} pipelines (Section~\ref{sec:background}), which will build the foundation for the subsequent steps in our audit pipeline.
In essence, for our implementation, we transform our audit pipeline into a \ac{cicd} pipeline.
We utilise the Tree-Sitter\footnote{Tree-Sitter: \url{https://tree-sitter.github.io/tree-sitter/} (accessed on 11.06.2024)} library, which converts the analysis code into an \ac{ast} as shown in Figure~\ref{fig:combined}.
Tree-Sitter has the advantage of being compatible with various common programming languages beyond Python.
The ASTs generated are then transformed into Turtle/RDF format and combined with additional metadata such as Train name or version as depicted in Figure~\ref{fig:combined}.
To manage the graph data, we utilise a blazegraph instance\footnote{Blazegraph: \url{https://blazegraph.com} https://tree-sitter.github.io/tree-sitter/}. 
This instance also serves as the backend for a custom web service.
This service enables researchers to access and inspect the artefacts stored in the graph database.

\paragraph{Static Code Analysis (Step 2-4)}

After the code has been transformed into an \ac{ast} and enriched with metadata, the subsequent phase involves identifying vulnerabilities within the code base.
In this particular Aggregation State 1, where the code is considered to be not executable, \ac{sast} becomes essential (see Section~\ref{sec:background}).
At this point, another advantage of our \ac{ast} approach mentioned above becomes evident:
Common \ac{sast} approaches utilise ASTs to examine the code structure and identify common patterns of vulnerabilities~\cite{yamaguchi2012SAST}.
Consequently, we can reuse the previously generated \ac{ast} for this \ac{sast} phase (Step 2).
Particularly in the context of our \ac{pht} scenario, vulnerabilities like injection flaws\footnote{Improper Control of Generation of Code ('Code Injection'): \url{https://cwe.mitre.org/data/definitions/94.html} (accessed on 11.06.2024)} could pose risks as they might lead to the injection of malicious code into Stations (see Table~\ref{tab:code_threats}).
Beyond the vulnerabilities of the code base directly, we additionally apply \textit{dependency scanning} (Step 3) to detect vulnerabilities introduced through external dependencies or libraries.
Ultimately, we integrate \textit{secret detection} (Step 4) into our process.
This step is important as the Train connects to a database (see Figure~\ref{fig:pht}) that holds sensitive data, hence requiring the use of credentials.
Exposed secrets in code repositories, especially in public or shared repositories, pose threats~\cite{dahlmanns2023secretsindocker}.
Therefore, credentials, such as passwords, \ac{api} keys, cryptographic keys, and other confidential data, should not be embedded within the source code of the Image/Train, e.g. as discussed in the work by Dahlmanns et al.~\cite{dahlmanns2023secretsindocker}.
Secret detection can prevent such vulnerabilities from being exploited, and developers can reduce the risk of data breaches and unauthorised access~\cite{dahlmanns2023secretsindocker}.

\textbf{Implementation.}
We rely on GitLab's built-in functionalities for the static code analysis, which can be reused and seamlessly integrated into the \ac{cicd} pipeline.
For the \ac{sast} aspect, we utilise GitLab's \ac{sast} Scanning\footnote{GitLab \ac{sast}: \url{https://docs.gitlab.com/ee/user/application\_security/sast/} (accessed on 11.06.2024)}, which encompasses both dependency scanning and secret detection.

\paragraph{Allow/Disallow Lists (Step 5)}

So far, we have considered vulnerabilities that are known and part of public databases.
While these vulnerabilities might still be relevant, they may not cover domain- or case-specific requirements.
Since \ac{pht} ecosystems can be applied to various domains with specialised requirements, there is a need for custom rule definitions.
Hence, we extend our pipeline with a customisable component, such as allow and disallow lists, for syntactical rules.
These allow or disallow lists bring a level of specificity and adaptability to our vulnerability checks, making them more flexible for the varied and specialised environments in which \ac{pht} ecosystems operate.

\textbf{Implementation.}
We use regular expressions to identify prohibited code patterns or commands within the code base.
This check uses a custom shell script designed to recognise and flag specific regular expressions in the code.
The regular expressions can be specified by the host of the \ac{pht} infrastructure.
Examples of regular expressions for, e.g. secrets, can be found in the work by Dahlmanns et al.~\cite{dahlmanns2023secretsindocker}.

\paragraph{Compliance Check (Step 6)}
While the allow/disallow component is tailored towards syntactical checks, we introduce a more advanced component that we refer to as a compliance check.
This component is designed to align with ecosystem-specific policies, for example.
Beyond checking the syntax of the code, the compliance check will ensure that the code adheres to certain predefined standards and rules that are specific to the particular ecosystem in which the Train is operating.
These standards can cover security, data handling practices, conventions, or regulatory requirements unique to the operational context.

\textbf{Implementation.}
We have set an exemplary policy that requires Python as programming language.
Any code written in a different language will not be accepted.
Additionally, we require the use of the PADME-Conductor library\footnote{PADME-Conductor: \url{https://pypi.org/project/padme-conductor/} (accessed on 11.06.2024)} within the Train code.
This library offers a range of predefined methods that are specifically designed to simplify the interactions between the Train and the Stations, such as data loading, analysis execution, result storage, and logging (similar to the steps in Figure~\ref{fig:pht}).
Setting the PADME-Conductor library as a compliance rule, we implicitly require that the code follows a certain structure that ensures consistency across different Trains and their transparency.
In general, this library can be considered secure and, thus, suitable for code compliance measures.
The concept of a secure library for the \ac{pht} has been inspired and suggested by the work of Wirth et al.~\cite{Wirth2021}.
These two compliance rules, Python and PADME-Conductor, are verified on an ad hoc basis using two shell scripts that scan the dependencies (e.g. the \texttt{Dockerfile} or the \texttt{requirements.txt}) of the Container Train for compliance.
Other implementations beyond ours might be possible to validate the compliance according to a policy.

\subsubsection{Detection in Aggregation State 2 (Encapsulation State)}

Based on the static code, as discussed in Section~\ref{subsec:requirementanalysis}, the code base is transformed into a software image, which is represented by Aggregation State 2.
In this phase, various software packages are merged to create a new software artefact, which may not be adequately scanned using the methods applied in Aggregation State 1.
Therefore, this particular Aggregation State 2 necessitates an approach different from static code analysis (Section~\ref{subsubsection:countermeasuresaggrstate1}) to detect vulnerabilities in the new image.
Various types of vulnerabilities can be introduced into software images for several reasons, including the use of insecure base images or configurations, unnecessary redundancies, or the usage of unofficial images~\cite{wist2021vulDocker,shu2017dockerhub,combe2016docker}.
The presence of vulnerabilities in images commonly found in official repositories has been previously identified and explored, as highlighted in the studies by Wist et al. and Shu et al.~\cite{wist2021vulDocker,shu2017dockerhub}.
Due to this evidence, we have chosen to extend our audit pipeline by incorporating a component (Step 7) dedicated to detecting vulnerabilities in images.

\textbf{Implementation.}
For our purposes, we have chosen Snyk to conduct software image scans.
This decision is primarily based on Snyk's compatibility with our GitLab \ac{cicd} pipeline mentioned above.
Furthermore, Snyk manages a database of known vulnerabilities, and it supports multiple programming languages.
To integrate Snyk into \ac{pasta}, we added the Snyk component into the GitLab \ac{cicd} pipeline definition.
Note that other tools like Clair or Trivy might also be possible.

\subsubsection{Detection in Aggregation State 3 (Execution State)}
\label{sec:detectionas3}
In the preceding sections, we discussed approaches primarily focused on analysing static and non-executed code.
However, Aggregation State 3 represents Trains in their executable form.
Therefore, it becomes necessary to identify problems that become apparent only when the Train is running.
One approach that can cover vulnerability detection during runtime is \ac{dast}, as introduced in Section~\ref{sec:background}.
Beyond vulnerabilities from a software perspective, \ac{dast} can also be used for benchmarking a Train and checking its efficiency regarding resource consumption.
Within the context of the \ac{pht}, \ac{dast} becomes relevant for multiple reasons.
For example, a malicious researcher might program Trains to intentionally communicate with external servers to extract raw data and transmit sensitive information to external and, therefore, unauthorised locations (for example, see Table~\ref{tab:code_threats}).
Additionally, \ac{dast} can identify Trains that are inefficiently programmed and consume excessive resources.
Such resource-intensive Trains can strain the Station’s infrastructure, potentially leading to performance degradation or even system failures. 
Further, \ac{dast} can provide transparency regarding changes in the content of a Train.
Essentially, it can reveal the data entering the Station and the data leaving it.
For instance, if the size of the Train increases significantly after execution, it could indicate that privacy-sensitive information has been collected, either intentionally or unintentionally.
\ac{dast} can make these changes in the Train content visible.

\textbf{Implementation.}
In general, we interpret the \ac{dast} process for a Train as the simulation of its execution.
A key challenge encountered in this simulation is ensuring the availability of data that the Train requires for its execution.
Therefore, we need either synthetic test data or, in an ideal scenario, actual sample data to conduct this simulation effectively.
Moreover, our \ac{dast} requires a container execution environment to run the Train code, as well as to simulate various Stations along its intended route.
In one of our previous works by Welten et al., we conceptualised and developed a simulation engine specifically designed for Trains~\cite{Welten2024,sven_weber_padme-phtplayground_2024}.
This engine can be leveraged to perform \ac{dast}, which involves simulating the operation of the Train and the Stations providing data that resemble real-world data.
This previous work provides interfaces that communicate with \ac{pasta}, as well as a framework for Train testing, benchmarking, and simulation to detect runtime issues.
In our process, the audit pipeline first uploads the Train code to the simulation engine, which builds the Train image.
Following this step, the Train is executed and benchmarked against several pre-defined and simulated Stations.
After the simulation, the engine returns key performance metrics, including CPU usage, I/O data, or content changes within the Train.
These performance metrics can then be used for an overall \ac{dast} assessment.
This simulation engine is also available as open-source and can be deployed complementary to \ac{pasta}~\cite{sven_weber_padme-phtplayground_2024}.

\subsubsection{Final Audit Decision}
After going through the individual steps of the security audit pipeline, the final step is to recommend whether or not the Train is accepted (Steps 9--10).
In the decision step, the results of the previous pipeline steps are fetched, and for each step, a score is calculated by counting the number of vulnerabilities.
Next, the scores are evaluated following a user-defined decision model.
In our implementation, this model is a decision tree.
Our assumption is that the host of the \ac{pht} infrastructure defines the conditions that must be met for a Train to pass.
Hence, administrators can adjust the acceptance criteria according to their own concerns, matching their respective use cases.
A decision model is defined by providing a threshold for each pipeline step (\ac{sast}, secret detection, image analysis, etc.) and a decision tree that determines which combination of thresholds has to be met for a positive decision.
For each pipeline step, the decision script compares the calculated score to the corresponding threshold and marks this step as \textit{passed} if the score is greater or equal to the provided threshold.
The host is free to choose subsets of thresholds as criteria.
Thus, the host is able to construct a custom logic by defining multiple branches that lead to acceptance.
An example of such a decision tree is given in the supplemental materials for reference~\cite{karlkindermann_saweltpasta-4-pht_2024}.

\begin{figure}[h]
    \includegraphics[width=\linewidth]{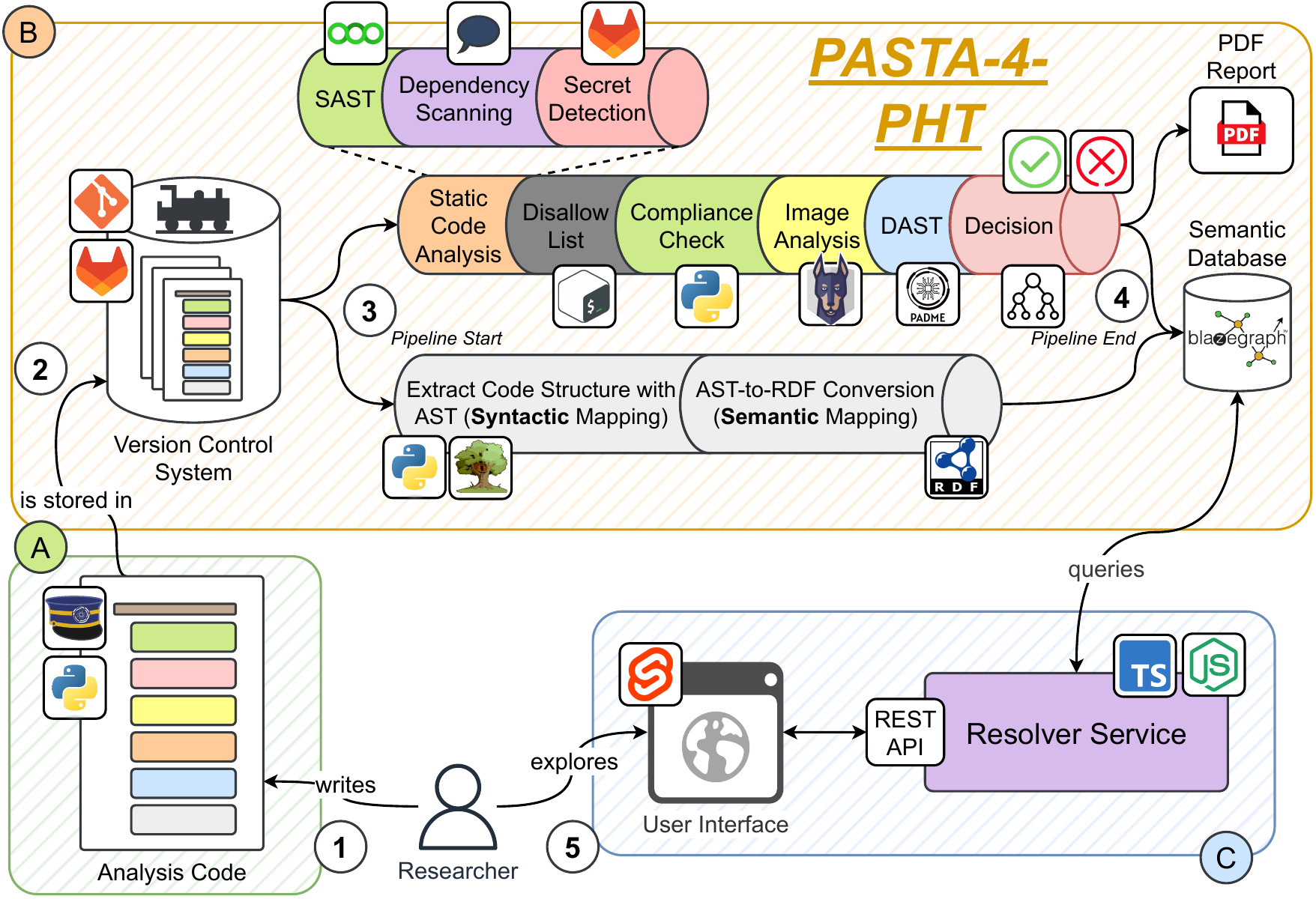}
    \caption{Overview of \ac{pasta}. In Step 1, the code is developed. After the upload to the \acf{vcs} (Component A), our audit pipeline is triggered that performs the steps we have defined in Figure~\ref{fig:formalPipeline}. After all steps have been performed, our audit pipeline (Component B) automatically produces a PDF report about potential vulnerabilities inside the Train, which can be used for decision-making. Further, for the sake of documentation, the Train code is stored in a graph database, which can be queried by the developer through a resolver service (Component C).}
    \label{fig:pasta4pht}
\end{figure}

As illustrated in Figure \ref{fig:pasta4pht}, the decision step provides more detailed information for each audit.
All audit-relevant assets are stored in the graph database (see Step 1), which can be used later to generate a more comprehensive view of the results in the user interface.
Note that all audit artefacts, such as specific vulnerabilities, violations or benchmark parameters, are managed and stored in a machine-readable format.
This facilitates their automatic incorporation into an audit report, such as a PDF document, representing the final output of \ac{pasta}.
This document can be used, for example, for the mandatory documentation of data processing activities (see Article 35 \ac{gdpr}).

Up to this point, our pipeline is ready for operation and we will evaluate it by using several trains in the next section.

\section{Results}
\label{sec:results}
In this section, we present the results of our evaluation of \ac{pasta}, which we categorised into a controlled setting (Section~\ref{subsection:invitro}) and real-world scenarios (Section~\ref{subsection:invivo}).
The controlled audit is inspired by the potential threat and attack surfaces detailed in Table~\ref{tab:code_threats} and may represent real attack scenarios.
Additionally, the real-world audit examines real-world Trains used in actual studies, making these Trains representative of applying \ac{pasta} in a practical and real-world situation.
Be aware that all Trains utilise Docker for containerising the Train code.
Additionally, note that our evaluation results do not incorporate the outcomes from allow and disallow lists or compliance checks, as these results can vary depending on the predefined criteria of what is considered permissible or not.
All audit outcomes, the PDF reports, and artefacts can be found in the supplementary materials~\cite{karlkindermann_saweltpasta-4-pht_2024}.

\subsection{Controlled Audit}
\label{subsection:invitro}
In the first part, we selected for each step one step-relevant vulnerability and created Trains for each selected vulnerability.
This has the purpose of validating that each step is conceptually working.
After we parsed the Train through \ac{pasta}, we check if it gets detected or not.
Note that we added the corresponding identifiers of the vulnerabilities to each of the following use cases.

\subsubsection{Use Case: \ac{sast}}
\paragraph{Introduced Vulnerability: Arbitrary Code Execution [bandit.B403]}
The \ac{sast} part of the pipeline is able to check the code inside the Train for vulnerabilities like injection flaws before executing it.
We created a Train that uses a so-called \textit{pickle} file\footnote{Pickle documentation: \url{https://docs.python.org/3.11/library/pickle.html} (accessed on 11.06.2024)} to load serialised objects from an external source, potentially allowing for arbitrary code execution.
The security pipeline successfully identified the deserialisation of untrusted data and included the vulnerability in the report.

\paragraph{Introduced Vulnerability: SQL Injection [bandit.B608]}
In addition to executing malicious code on the Station, we also want to prevent the injection of special elements in a database query, such as \textit{SQL injection}.
While some databases allow for a restriction of commands that will be executed, e.g. disabling \textit{DELETE} or \textit{UPDATE}, as they would compromise the integrity of the data, we generally want to prevent users from injecting unauthorised queries in SQL commands.
The SQL injection was discovered within the \ac{sast} state of the security pipeline.

\subsubsection{Use Case: Dependency Scanning}

\paragraph{Introduced Vulnerability: Vulnerable Python Library [\ac{cve}-2018-18074]}
We also want to identify vulnerabilities that have not been introduced deliberately. An example of such a vulnerability would be the inclusion of vulnerable libraries or, more generally, all kinds of dependencies on third projects containing exploitable vulnerabilities.
We introduced a Train installing version \textit{2.18.4} of the \textit{requests} library for Python, which is known for having several critical vulnerabilities.
The security pipeline successfully identified and reported these vulnerabilities in Step 3 (see Section~\ref{subsection:methodolody}).

\subsubsection{Use Case: Secret Detection}

\paragraph{Introduced Vulnerability: \ac{ssh} Key [bandit.B105]}
Another common vulnerability issue arises from using unencrypted credentials, which might be stored in plaintext within the code file. Secret detection scans for secrets matching a certain pattern, including but not limited to \ac{ssh}-keys, PGP-keys and credentials to cloud service providers like \textit{Amazon Web Services (AWS)}.
Step 4 successfully identified the Train as vulnerable because it stored a string of an \ac{ssh} private key in plaintext.

\subsubsection{Use Case: Image Analysis}

\paragraph{Introduced Vulnerability: Vulnerable Base Image [Debian:10]}
To test the image analysis (Aggregation State 2) part of the pipeline, we introduced a new Train using \textit{python:3.10.0a7-buster} as the base image.
This image uses \textit{Debian 10}\footnote{SNYK Results for Debian 10: \url{https://snyk.io/test/docker/debian:10} (accessed on 11.06.2024)} and an early developer version of Python 3.10 and is reported to have a particularly high number of vulnerabilities. One example of such a vulnerability within Debian 10 is [\ac{cve}-2022-23219], which poses the risk of a \ac{dos} attack utilising a buffer overflow.
The audit produced by the security pipeline successfully discovered this in Step 7, among other vulnerabilities listed in the vulnerability databases of Snyk or the \ac{cve} reference system.

\subsubsection{Use Case: \ac{dast}}
\paragraph{Introduced Vulnerability: Network Communication}
Network communication can risk the confidentiality of the Station's data, as sensitive information could be transmitted to a third party during runtime.
Similarly, incoming network traffic would enable users to load and execute arbitrary code on runtime, which is why it is crucial to detect both.
To demonstrate the monitoring of network traffic, we created a Train that initiates a local database connection before sending a POST request to an outside server.
Finally, the Train retrieves data from a second outside host via a GET request.
Using the \ac{dast} component (see Section~\ref{sec:detectionas3}), the security pipeline detected 56.1 kB of incoming traffic and 14.7 kB of outgoing traffic.
This demonstrates that the \ac{dast} service we are using is able to make (unwanted) network communication visible.
Note that this vulnerability does not have a specific identifier, as our goal is to detect network traffic in general.

\subsection{Real-world Audit}
\label{subsection:invivo}

After we have tested our audit pipeline in a controlled setting by deliberately introducing various vulnerabilities, we now intend to evaluate the pipeline in a real-world context.
We aim to apply the pipeline to actual Trains used in prior \ac{pht} studies about Breast Cancer, Skin Lesion, and Basic Query Train that we have conducted in the past~\cite{weltenMultiInstitutionalBreastCancer2022, mouDistributedSkinLesion2021, Welten2024}.
Overall, we want to audit five trains of different complexities and purposes, ranging from basic patient counting to Machine Learning and different sizes.
An overview is given in Table~\ref{tab:vivoTrains}.

\begin{table}[htbp]
\centering
\caption{Comparison of tasks, data types, and analysis methods across various Trains that are used to validate \ac{pasta}.}
\label{tab:vivoTrains}
\begin{tabular}{|c| c | c | c | c | c|} 
 \toprule
 & \makecell{Basic Query \\ Train ~\cite{Welten2024}} & \makecell{ISIC \\2019 I \cite{mouDistributedSkinLesion2021}} & \makecell{ISIC \\2019 II \cite{mouDistributedSkinLesion2021}} & \makecell{Breast \\Cancer I \cite{weltenMultiInstitutionalBreastCancer2022}} & \makecell{Breast \\Cancer II \cite{weltenMultiInstitutionalBreastCancer2022}} \\
 \midrule
 Task & \makecell{Counting \\ patients} & \makecell{Gather\\ statistics} & \makecell{Train\\ model} & \makecell{Gather\\ statistics} & \makecell{Train\\ model} \\ 
 \hline
 \makecell{Data} & \makecell{Relational\\ Database} & \makecell{FHIR,\\ Images} & \makecell{FHIR, \\ Images} & \makecell{File Dump\\ .CSV} &\makecell{File Dump\\ .CSV} \\ 
 \hline
 Analysis & \makecell{SQL Count\\ Query} & \makecell{Statistical\\ Libraries} & \makecell{ResNet-18 \\(PyTorch)} & \makecell{Statistical\\ Libraries}& \makecell{Log. Reg. \&\\ GAN \\ (PyTorch)} \\
 \hline
 \makecell{Analysis\\ Results}& \makecell{.TXT} & \makecell{Plots \& .CSV} & \makecell{.TAR\\ (Model)} & \makecell{.TXT} & \makecell{.TAR\\ (Model)} \\ 
 \hline
 \makecell{Lines of\\ Code} & 47 & 220 & 460 & 44 & 696 \\
 \hline
 Image Size & 1.02 GB & 1.22 GB & 5.65 GB & 1.02 GB & 7.07 GB \\
 \hline
 \makecell{Number of\\ Dependencies}& 2 & 6 & 12 & 2 & 13 \\
 \bottomrule
\end{tabular}
\end{table}

As outlined in our workflow shown in Figure~\ref{fig:pasta4pht}, the process begins with the uploading of the Train code and its necessary build files to the GitLab repository (the \ac{vcs} in Figure~\ref{fig:pasta4pht}).
Once the code is pushed to the repository, the audit pipeline is automatically activated, which processes the source code.
We discuss the results in the following section.

\subsubsection{Outcomes}
\label{sec:outcomes}

We summarised the results of the real-world audits in Table~\ref{tab:securityperformance}.
For a detailed overview of the audit results, we refer to the supplementary materials, where we provided the PDF reports of each Train~\cite{karlkindermann_saweltpasta-4-pht_2024}.
These reports also include the Train code, the descriptions of the vulnerabilities given in Table~\ref{tab:securityperformance} and their location in the code, a file tree of the changed files within the Train during the \ac{dast}, and a final decision about the Train's approval.
Be aware that the decision taken was based on thresholds that were set arbitrarily.

\begin{table}[htbp]
\centering
\caption{Results of our real-world audit. The static code analysis classifies the vulnerabilities into Low/Medium/High/Critical. The \ac{dast} produces benchmarks in terms of CPU, memory, number of threads, sending (TX), and reading (RX) operations. Information about the selected \ac{dast} metrics can be found here\protect\footnotemark. We consider lower benchmark metrics to be preferable.}
\label{tab:securityperformance}
\begin{tabular}{|c| c | c | c | c | c|}
 \toprule
 & \makecell{Basic Query \\ Train ~\cite{Welten2024}} & \makecell{ISIC \\2019 I \cite{mouDistributedSkinLesion2021}} & \makecell{ISIC \\2019 II \cite{mouDistributedSkinLesion2021}} & \makecell{Breast \\Cancer I \cite{weltenMultiInstitutionalBreastCancer2022}} & \makecell{Breast \\Cancer II \cite{weltenMultiInstitutionalBreastCancer2022}} \\
    \midrule
    \makecell{\ac{sast}\\ (L/M/H/C)} & (0/1/1/0)  & (0/0/0/0) & (0/0/0/0) &(0/0/0/0) & (0/0/0/0)\\
    \hline
    \makecell{Secret\\ Detection\\ (L/M/H/C)} & (0/0/0/0) & (0/0/0/0) & (0/0/0/0)& (0/0/0/0) & (0/0/0/0)\\
    \hline
    \makecell{Dep.\\ Analysis\\ (L/M/H/C)} & (0/0/0/0) & (0/0/0/0) &(0/0/0/0) & (0/0/0/0) &  (0/0/0/0)\\
    \hline
    \makecell{Image\\ Analysis\\ (L/M/H/C)} & \makecell{181\\3\\1\\1} & \makecell{297\\168\\113\\35} & \makecell{22\\8\\0\\0}& \makecell{297\\168\\113\\35} & \makecell{297\\168\\113\\35} \\
    \hline
    \makecell{CPU\\Mem. in GB\\PIDs} & all $\leq.01$ &\makecell{4.0\\ ~0.13\\ 2} & \makecell{5.83\\ ~0.523\\ 11} & \makecell{$\leq.01$\\ ~0.031\\ 1} & \makecell{50.0\\ ~0.296\\ 30} \\
    \hline
    \makecell{RX in MB\\TX in MB}& all $\leq.01$ & \makecell{7.475\\0.236} & \makecell{45.261\\0.328} & \makecell{2.253\\0.051}  & \makecell{12.288\\0.276} \\
\bottomrule
\end{tabular}
\end{table}
\footnotetext{Docker Stats: \url{https://docs.docker.com/config/containers/runmetrics/} (accessed on 11.06.2024)}

Based on the statistics gathered from our audit pipeline, we derive the following observations.
The ISIC 2019 I, ISIC 2019 II, Breast Cancer I, and Breast Cancer II trains have no vulnerabilities detected by \ac{sast}.
The Basic Query Train exhibits low to medium-level vulnerabilities, which might indicate some areas of potential risk.
All trains successfully passed secret detection without any vulnerabilities, which shows that no data like passwords or \ac{api} keys are exposed in the code.
Similar to secret detection, all trains showed no vulnerabilities in their dependencies.
All Train images have vulnerabilities ranging from low to critical, with ISIC 2019 I and both Breast Cancer Trains having the highest severity.
This suggests a need for enhancement of the image's security.
Auditing real-world trains showed that analysing base images is particularly important, as it discovered several critical vulnerabilities in the used Linux distribution (Debian 10).
The used base image contains numerous vulnerabilities rated with a \ac{cvss} score of 9.0 or higher.
For example, among the critical vulnerabilities, we identified overflow-related issues (e.g. SNYK-DEBIAN11-AOM-1300249), misconfigurations (e.g. SNYK-DEBIAN11-CURL-2936229), and OS/SQL command injection flaws (e.g. SNYK-DEBIAN11-OPENSSL-2807596, SNYK-DEBIAN11-OPENLDAP-2808413).
While overflow vulnerabilities might not directly compromise data security, they could affect the Train's reliability, potentially leading to \ac{dos} attacks on the Station.
Injection vulnerabilities pose risks for data breaches or poisoning.
Further, the Basic Query Train and Breast Cancer I show minimal resource usage, indicating efficient performance.
The ISIC 2019 II and Breast Cancer II trains, however, exhibit higher CPU and memory usage, which suggests more intensive processing.
The number of processes (PIDs) also varies, which indicates differences in the number of processes or threads running for each Train.
The network I/O shows varied levels of network activity across the Trains, with ISIC 2019 II having the highest network usage - potentially due to data-intensive/reading (RX) operations.
Overall, the notable difference in image analysis vulnerabilities compared to other checks emphasises the importance of securing the images.

Eventually, our experimental audits demonstrate that our pipeline can effectively detect vulnerabilities.
According to the Snyk database, various fixes are available to mitigate these risks, allowing developers to update the base image and close these security gaps before the actual experiment in the \ac{pht} infrastructure.
Furthermore, the performance metrics show different levels of efficiencies among the trains, which could be due to their inherent complexities or potentially inefficient implementations.

\section{Discussion}
\label{sec:discussion}

In this section, we will discuss the results of our evaluation.
Initially, we discuss the results of our security audits and how they relate to governmental procedures and policies (Section~\ref{subsec:securityaudit}).
Following that, we outline our contributions to the \ac{fair4rs} (Section~\ref{subsec:contributiontoFAIR}).
Lastly, we address potential limitations of \ac{pasta} and our evaluation (Section~\ref{subsec:threatstovalidity}).

\subsection{Security Audit}
\label{subsec:securityaudit}

In the previous section, we demonstrated that \ac{pasta} is conceptually functional and successfully identifies a range of intentionally introduced vulnerabilities and vulnerabilities in Trains from previous studies.
In general, the final audit report reveals the first indicators for code improvements to enhance the overall quality and security of the Trains.
For example, our evaluation, as detailed in Section~\ref{sec:outcomes}, has demonstrated that the most significant source of vulnerabilities within the Trains stems from the Train image itself.
From a Train developer's perspective (e.g. a data scientist), \ac{pasta} scans the code before submission, whether it contains potential vulnerabilities.
The audit reports highlight such vulnerabilities in the Train images, providing developers with the information needed to select more secure base images.
For example, the \texttt{python:3.10.0a5} image\footnote{Snyk Results for Python 3.10.0a5: \url{https://snyk.io/test/docker/python\%3A3.10.0a5} (accessed on 11.06.2024)} is vulnerable to multiple critical severity issues.
Replacing it with the \texttt{python:3.12.0a5-slim} base image\footnote{Snyk Results for Python 3.10.0a5-slim: \url{https://snyk.io/test/docker/python\%3A3.10.0a5-slim} (accessed on 11.06.2024)} can reduce vulnerabilities while maintaining functionality, thereby decreasing the number of (potential) attack surfaces.

However, in Table~\ref{tab:securityperformance}, we see that vulnerabilities can increase rapidly.
This observation raises doubts about their risk to the Train software, especially considering false positives, false negatives, and relevance in the \ac{pht} context.
As we do not include a filtering mechanism in our audit pipeline, such filtering (relevance vs.\ non-relevance) has to be conducted manually, and we leave the question about relevance open for future work.
Also relevant to this discussion is identifying which actor can exploit these vulnerabilities, as this determines their relevance.
Our focus so far has been on attack vectors involving the Train creator.
Malicious Station admins are also considerable.
From a Station admin's perspective, one additional attack vector might be the input of non-sanitised data (into the Train) from a connected source that is passed to vulnerable code, potentially causing a data breach.
However, since \ac{dast} is included in our pipeline, such activities should be detected during the audit phase and the chance of a Station admin intentionally sabotaging their own Station (causing a data breach) is at least debatable.

In general, \ac{pasta} should not be understood as a holistic security auditing tool that covers every attack vector.
Instead, we interpret our work as support in two ways.
First, \ac{pasta} should be seen as a decision-making tool for Station admins and infrastructure hosts, who can review the audit and decide whether to execute a Train, ensuring each Train is checked before deployment.
Second, our work supports managing regulatory obstacles, particularly those presented by the \ac{gdpr}, along with the associated documentation requirements in a (research) project.
Given that the \ac{pht} is an enabler for conducting medical data science in research involving sensitive data, its usage and application inherently falls under the purview of \ac{gdpr} regulations. 
According to Article 35 of the \ac{gdpr}, the deployment of new technologies for data processing requires a so-called \ac{dpia}\footnote{A template is available at \url{https://gdpr.eu/data-protection-impact-assessment-template/} (accessed on 11.06.2024)}~\cite{chassang2017DPIA,bieker2016dpia}.
A \ac{dpia} is a component of the data protection framework that ensures that privacy and data protection are part of the operational practices of organisations, such as hospitals, which handle patient data~\cite{bieker2016dpia}.
Nonetheless, working with data protection regulations can often present challenges to researchers and other project personnel~\cite{sirur2018gdpr}.
Researchers must outline how data is processed, identify potential risks, propose mitigation strategies, and verify compliance with established data protection standards.
Using our audit pipeline that automatically generates \ac{dpia}-relevant assets has several advantages in that context.
First of all, it reduces the manual effort required in documenting the data processing activities and the assessment of data protection risks.
Since our pipeline systematically audits the Train code, the documentation process is consistent across different projects or studies within the same organisation.
It also minimises human errors that could occur in manually compiling \ac{dpia} reports, and we argue that once it has been installed, personnel with limited technical expertise can resort to our tool and create reports.
As the requirements for data processing may evolve or modifications to the Train code have been applied, our pipeline can be adjusted or re-executed accordingly.
This adaptability allows for a compliance assessment with changing conditions or regulations in a more timely manner.
Due to these reasons, we argue that our contribution fuels the management of current processes related to governance, data protection, and documentation of research assets as our tool circumvents the necessity for manual review of data analyses and the manual compilation of \ac{dpia} reports.

\subsection{Contribution to FAIRness in Research Software}
\label{subsec:contributiontoFAIR}

\begin{figure}
    \includegraphics[width=0.8\linewidth]{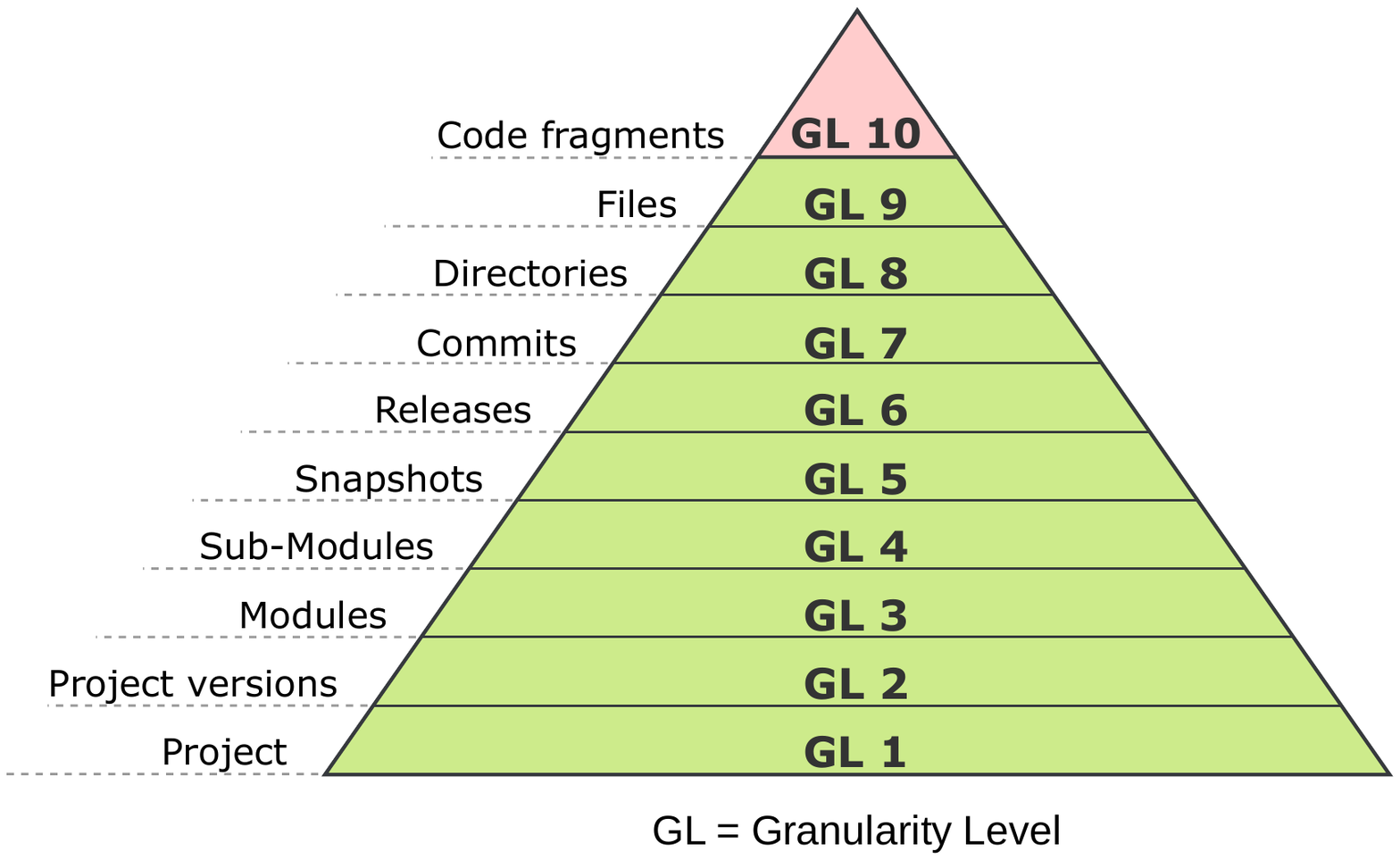}
    \caption{
        The \acf{gl} according to the \acf{fair4rs} working group.
        Starting at the coarsest level on the bottom (Project level \ac{gl} 1) to the finest at the top (Code fragments \ac{gl} 10).
        We inverted the order to reflect the \ac{gl} becoming finer as the pyramid sharpens. Adapted from the \ac{fair4rs}/\acf{rda}~\cite{RDAFORCE112020CodeID}.
    }
    \label{fig:Pyramid}
\end{figure}

In this section, we revisit the documentation aspect that has been introduced in the previous section.
Beyond the audit report, we also generate (meta)data in a graph format that we store in our semantic database (see Figure~\ref{fig:pasta4pht}).
This leads us into the broader context of the \ac{fair} Principles that we shortly introduced in Section~\ref{sec:background}.
However, we do not process research data per se but research software, which is represented by the Train and differs from data instances.
Research software, as defined by Hong et al., is software that is created during the research process or for a research purpose~\cite{ChueHong2022FAIR4RS}.
Software can encompass source code files, algorithms, scripts, computational workflows, and executables~\cite{hasselbringFAIRResearchData2020}.
In recent years, the FAIRification of research software has been recognised as similarly important for research because it also produces insightful data \cite{lamprechtFAIRPrinciplesResearch2020}.
However, since the application of \ac{fair} Principles to research software is comparably new, it has not yet reached the same research maturity as \ac{fair} data \cite{lamprechtFAIRPrinciplesResearch2020, ChueHong2022FAIR4RS,Barker2022FAIRSoftware}.
Therefore, the \ac{rda} has founded a working group to formulate the \ac{fair4rs}, which adapted the original \ac{fair} principles~\cite{ChueHong2022FAIR4RS}.
The adjusted \ac{fair4rs} can be applied to different levels of the software \cite{RDAFORCE112020CodeID,ChueHong2022FAIR4RS}.
These different levels were defined by a working group of the \ac{rda} as the different \ac{gl} of software (see Figure \ref{fig:Pyramid})~\cite{RDAFORCE112020CodeID}.
We can see the coarsest level starting at \textit{\ac{gl} 1 Project} and ending with the finest at \textit{\ac{gl} 10 Code fragments}.
Being able to reference the finest \ac{gl} 10 enables us to reference finer aspects of the software, for example, a function or an assignment of a variable.

Setting these layers in relation to our concept, we find that using the \ac{vcs} addresses the initial nine layers.
Complementing this, our approach using \acp{ast} for the semantic mapping gives a first foundation for Layer 10.
In a broader sense, our approach based on Semantic Web principles gives the necessary interfaces for integrating additional ontologies to enrich the Train code with more metadata.
One practical example concerning this use case at hand is the data linkage between the Train code and the detected vulnerabilities that are categorised and maintained by the initiatives presented in Section~\ref{sec:background}.
Recommendations or solutions to resolve vulnerabilities can also be linked to the corresponding detected vulnerabilities.
The graph-based structure further allows for meta-analysis of the code to identify recurring patterns (such as identical lines of code) or common vulnerabilities between Trains, as well as tracking the code's provenance and evolution of code to understand its history and modifications over time.
Hence, our approach lays the groundwork towards the overarching aim of fostering maximum transparency and FAIRness in the Train development and deployment processes.

\subsection{Limitations}
\label{subsec:threatstovalidity}

Regarding the limitations, we want to point out that our approach does not guarantee holistic security coverage.
Specifically, vulnerabilities introduced by the \ac{pht} infrastructure were not within the scope of our work.
For instance, we did not conduct checks to determine if a Station manipulates the Train.
However, the work of Herr et al., which introduces concepts for manipulation detection during Train transmission, can be seen as complementary to ours~\cite{herr2022bringing}.
Another significant challenge is the effectiveness and relevance of detected vulnerabilities:
As mentioned in Section~\ref{subsec:contributiontoFAIR}, the effectiveness of our pipeline partially relies on the effectiveness of third-party tools (e.g. the image analysis tool).
It might be the case that some tools may work better, or others may not detect all vulnerabilities. 
Some related works already compared common tools in this regard~\cite{tunde-onadele2019,javed2021}.
While our pipeline identifies numerous vulnerabilities within, e.g. the software images, filtering the vulnerabilities according to their relevance remains also an open question and challenging.
This limitation aligns with the findings of Rajapakse et al., who highlighted the challenge of a high number of false positives in security audits and DevSecOps~\cite{challengesindevsecops}.
Another issue is the audit granularity, particularly in the context of network I/O monitoring during \ac{dast}.
We have demonstrated the capability to detect network I/O activities (see Table~\ref{tab:securityperformance}).
This detection mechanism may become biased when the Container Train accesses actual data, as these legitimate data queries are also classified as network traffic if no filtering mechanism is applied.
As a result, distinguishing between malicious and legitimate network activity becomes challenging, rendering our network I/O assessments less precise.
A possibly more effective and complementary approach could involve isolating the Train execution in a sandbox environment within the Station to prevent external communication instead of ad hoc audits prior to the deployment.
In relation to \ac{dast}, our pipeline's execution coverage is currently limited by the \ac{dast} component.
A stronger attacker model could force the \ac{dast} component to verify a clean programming path during the audit phase while the production phase introduces a malicious one.
Expanding the coverage to encompass all programming paths could be considered for future enhancements.
Further, determining an appropriate approval threshold for the Train approval also presents a challenge.
The set threshold in our study is somewhat arbitrary, highlighting the need for a more systematic method to assess what level of risk is acceptable in the context of the \ac{pht}.
Despite these limitations, \ac{pasta} offers an initial assessment of the Train's security that can be considered in an approval committee.
Our work lays the groundwork for enhancing the security of \ac{pht} applications and offers the essential artifacts for setting up such a pipeline, which can be accessed online~\cite{karlkindermann_saweltpasta-4-pht_2024}.
\ac{pasta} is automated, adheres to best practices, and can serve as inspiration for additional pipeline components.
Hence, we argue that any future use of the \ac{pht} can be appropriately secured by utilising and relying on \ac{pasta}, our pipeline for automated security and technical audits.

\section{Conclusion}
\label{sec:conclusion}

In this study, we proposed \ac{pasta}, a tool for detecting vulnerabilities within the \ac{pht}.
Based on the different states of a Train along its lifecycle, we defined state-specific approaches to detect vulnerabilities in the code.
We combined all approaches for each Aggregation State into a software pipeline called \ac{pasta}.
Our reference implementation, which is publicly available, draws inspiration from DevSecOps pipelines and produces audit reports fully automated.
We demonstrated that our pipeline can expose deficiencies in Trains and support the development process.

In general, the broader impact of our tool lies in fostering a more secure environment for distributed health data analytics, thereby promoting trust and reliability in applications like the \ac{pht}.
By automatically scanning for vulnerabilities, our approach enhances the security of the \ac{pht} ecosystem beyond the current state-of-the-art.
This contributes to the acceptance and adoption of the \ac{pht} framework.
Furthermore, our work bridges the gap between security assessments of the \ac{pht} and the requirements of current data protection frameworks and governance needs.
By streamlining the process of identifying and documenting vulnerabilities, our pipeline provides automated assistance to researchers with their documentation needs, such as DPIAs.
This automation reduces regulatory hurdles and simplifies both the initial preparation and the overall conduct of data analysis on sensitive data, tasks that traditionally had to be done manually.

Looking beyond the initial maturity of \ac{pasta}, we have identified several areas for future work in both implementation and research.
Currently, we use our own audit report template.
Future efforts could focus on generating PDF reports automatically from our pipeline's output, ensuring they align with prevalent \ac{dpia} policies or templates.
Automating this process has the potential to accelerate clinical research by reducing the manual overhead involved in producing necessary governance documentation.
Furthermore, our work establishes a foundation for advanced vulnerability or security scanning methods that could be explored in future research beyond our proposed concept.
For instance, utilising semantic contexts derived from the analysis code, future research could involve a meta-analysis of the code base.
This meta-analysis could cover aspects such as detecting similarities or supporting the definition of standard code components considered free from vulnerabilities.
It would also involve assessing the relevance of a vulnerability within the specific \ac{pht} context and determining its exploitability, thereby establishing whether the identified vulnerability poses a real threat or can be considered a false positive.
However, our available data is insufficient to make precise statements on the exploitability of identified vulnerabilities, necessitating more efforts in that direction. A more comprehensive dataset (including more Trains) and further analysis would be required to accurately assess the practical risks associated with these vulnerabilities within the \ac{pht} framework.
We look forward to corresponding future developments and are open to receiving (code) contributions for our artefacts.

\backmatter



\bmhead{Acknowledgements}
The authors thank Sven Weber for his valuable feedback during the pipeline development.
\section*{Declarations}

\begin{itemize}
\item Funding: No funding received.
\item Conflict of interest/Competing interests (check journal-specific guidelines for which heading to use): The authors declare that they have no competing interests.
\item Ethics approval and consent to participate: Not applicable.
\item Consent for publication: Not applicable.
\item Data, Materials, and Code availability:
We have uploaded all relevant artefacts related to our study to Zenodo~\cite{karlkindermann_saweltpasta-4-pht_2024}. The repository can be found at the following DOI: \url{https://doi.org/10.5281/zenodo.11505228}.
This repository contains the Train code used, audit reports, pipeline step codes, and screenshots of the final audit reports.
\item Author contribution:
S.W. conceptualised the work, supervised the sub-component development, conceptualised the manuscript, and wrote it.
K.K. reviewed and wrote the manuscript, contributed to all aspects of the pipeline, and worked on the frontend component.
A.P. reviewed this work, contributed to the security aspects, and worked on integrating the \ac{sast} and image security audits.
M.G. reviewed this work, contributed to the semantic component, also contributed to the integration of the \ac{dast} component, and worked on the frontend component.
M.J. reviewed this work and contributed to the security aspects.
L.N. worked on the semantic integration with the existing metadata schema and reviewed the work.
A.N., J.L., and J.P. reviewed this work.
S.D. supervised the work and supported the semantic integration.
\end{itemize}

\printacronyms

\bibliography{sn-bibliography}

\end{document}